\documentstyle[twocolumn,epsf,axodraw,amssymb]{article}
\def\section#1{\bigskip\centerline{\bf #1}\medskip}

\begin{document}

\title{
       \rightline{\small quant-ph/0103017}
{\Large\bf Carbon---the first frontier of information processing}}

\author{\normalsize APOORVA PATEL\\
        \normalsize\it
        CTS and SERC, Indian Institute of Science, Bangalore-560012, India\\
        \normalsize\it
        (Fax, +91-80-3600106; E-mail, adpatel@cts.iisc.ernet.in)}

\date{\small
Information is often encoded as an aperiodic chain of building
blocks.\hfill Modern digital computers use bits as the building\\
blocks, but in general the choice of building blocks depends on
the nature of the information to be encoded.\hfill What are the\\
optimal building blocks to encode structural information?\hfill
This can be analysed by substituting the operations of addition\\
and multiplication of conventional arithmetic with translation and
rotation.\hfill It is argued that at the molecular level, the best\\
component for encoding discretised structural information is
carbon.\hfill Living organisms discovered this billions of years ago,\\
and used carbon as the back-bone for constructing proteins that
function according to their structure.\hfill Structural analysis\\
of polypeptide chains shows that an efficient and versatile
structural language of 20 building blocks is needed to implement\\
all the tasks carried out by proteins.\hfill Properties of amino
acids indicate that the present triplet genetic code was preceded\\
\leftline{by a more primitive one, coding for 10 amino acids using
          two nucleotide bases.}
\medskip
\leftline{{\bf Keywords.}
Amino acid; aminoacyl-tRNA synthetase; computation; genetic code; information;
lattice models; protein}
\leftline{structure; quantum search; tetrahedral geometry}
\medskip
\leftline{\small [Patel A 2002 Carbon---the first frontier of information
processing; {\it J. Biosci.} {\bf 27} 207-218]}
\medskip\hrule}

\maketitle

\section{1. Structural information}

It is a characteristic of living organisms to acquire information,
interpret it and pass it on, often using it and refining it along the way.
This information can be in various forms or languages. It can be genetic
information passed on from the parent to the offspring, sensory information
conveyed by the sense organ to the brain, linguistic information communicated
by one being to another, or numerical data entered in a computer for later use.
It is advantageous to process the information efficiently, and not in any
haphazard manner. In case of living organisms, Darwinian selection during
evolution can be considered the driving force for such optimisation. In
general, information processing is optimised following two guidelines:
minimisation of physical resources (time as well as space), and minimisation
of errors.

A striking feature of all the forms of information listed above is that the
messages are represented as aperiodic chains of discrete building blocks.
Such a representation, called digitisation of the message, is commonplace
due to its many advantages. Discretisation makes it possible to correct
errors arising from local disturbances, and so it is desirable even when the
underlying physical variables are continuous (e.g. voltages and currents in
computers). It is also easier to handle several variables each spanning a
small range than a single variable covering a large range. Any desired
message can then be constructed by putting together as many as necessary of
the smaller range variables, while the instruction set required to manipulate
each variable is substantially simplified. This simplification means that
only a limited number of processes have to be physically implemented leading
to high speed computation. An important question, therefore, is to figure out
the best way of digitising a message, i.e. what should be selected as the
building blocks of the aperiodic chain.

The information contained in a message depends on the values and locations
of the building blocks. Given a set of building blocks, Shannon quantified
the information contained in a message as its entropy, i.e. a measure of
the number of possible forms the message could have taken. This measure
tells us that the information content of a message can be increased by
eliminating correlations from it and making it more random. It also tells
us that local errors in a message can be corrected by building long range
correlations into it. But it does not tell us what building blocks are
appropriate for a particular message. The choice of building blocks depends
on the type of the information and not on the amount of information.

Information can be translated from one language into another by replacing
one set of building blocks used to encode the information by another, e.g.
textual information is stored in the computer in a binary form using the
ascii code. Nonetheless, physical principles are involved in selecting
different building blocks for different information processing tasks.
For example, our electronic computers compute using electrical signals but
store the results on the disk using magnetic signals; the former realisation
is suitable for quick processing while the latter is suitable for long term
storage. In selection of building blocks with appropriate properties, the
foremost practical criterion is that it should be easy to distinguish one
building block from another. This simple criterion allows control over
errors, and is often sufficient for a heuristic understanding of the number
of building blocks of our languages. We use decimal system of numbers
because we learnt to count with our fingers. The number of phonemes in our
languages (i.e. vowels and consonants but not the tone) are determined by
the number of distinct sounds our vocal chords can make. Computers and
nervous systems use binary code because off/on states can be quickly
decided with electrical signals. Genetic information is encoded using four
nucleotide bases, perhaps because the quantum assembly algorithm is the
optimal choice for replication at the molecular scale (Patel 2001).

Numerical representation of information is one-dimensional and uses building
blocks with an ordering amongst them (e.g. one is greater than zero). But
these features may not be present in other types of information. For example,
ordering is not required for letters of an alphabet, and representation of
structural information requires higher dimensional building blocks. To find
the building blocks most suitable to encode a particular type of information,
one has to closely inspect the relation between the type of information and
the physical properties and tasks associated with it.

The information in the genes for the synthesis of proteins is a clear-cut
example of structural information. Living organisms generally do not have
access to desired biomolecules in a readymade form. They first break down
the ingested food into small building blocks, and then assemble the pieces
in a precise manner to synthesise the desired biomolecules. The hereditary
DNA is a one-dimensional read-only-memory in this process; the original DNA
strand remains unchanged while the information contained in it is copied on
the new strand that is assembled on top of it. The genes carry the blueprint
of how to synthesise proteins by joining together their building blocks---the
amino acids. The role of a protein in biochemical reactions is determined by
its three-dimensional shape and size, and the precise arrangement of chemical
groups at its reaction sites. This three-dimensional structural information
of proteins is encoded as a one-dimensional chain of amino acids, with the
interactions amongst the amino acids determining how the chain would bend
and fold to produce protein structures. To investigate the details of this
mechanism, it is natural to ask: what is the best way of encoding structural
information? This is the question addressed in this work.

Many well-known properties of proteins and their relation to structural
information are summarised in the Appendix. This material is provided
only for quick reference, and those familiar with it can easily skip it.
Every protein does not necessarily display all these properties, and
there is considerable variation in the behaviour of different proteins
(e.g. between small and large proteins). On the other hand, it has to be
emphasised that an efficient and yet versatile langauge must be capable of
incorporating all the desired features, even though specific instances of
the language may not display every possible feature. The ideal language
for proteins, therefore, must have the capability to: (1) fold linear
chains into three-dimensional structures and also unfold them, (2) form
three-dimensional structures of different shapes and sizes, (3) include
different chemical groups as part of the amino acids, and (4) show chiral
behaviour.

Any structural transformation of a rigid body can be described in terms of
two basic operations, translations and rotations. The set of all rigid body
translations and rotations forms the well-known Galilean group, which has
been studied in detail by physicists. To construct the building blocks of
structural information, we have to discretise this continuous group and yet
maintain its features required to encode information.

We can compare translations and rotations to the fundamental operations of
arithmetic---addition and multiplication. While addition is nothing but
translation along the real line, multiplication is quite different from
rotation. Rotations in our three-dimensional space are not commutative
and that is of crucial importance in representing structural information.
(The group of three-dimensional rotations is $SU(2)$, which can be
represented using Pauli matrices or quaternions.) The building blocks of
numerical information are elements of $Z_n$, the group of integers modulo
$n$, and the cyclic nature of this group represents the order amongst the
building blocks. The building blocks of structural information need to have
characteristics of rigid bodies, i.e. specific size and orientation in
three-dimensional space. To find them we have to look for a finite
non-commutative group. In addition, to address the question of protein
structure, we should look for transformations that take place at the atomic
scale.

Translations are easily discretised, as uniformly spaced units along a
polymer chain. The atomic structure of matter provides a natural unit for
translation---the physical size of the building blocks. Indeed, the amino
acids making up proteins differ from each other in terms of their side
chemical groups, while their components along the chain are identical.
Any translation can be built up from the elementary operations of addition
of a building block, deletion of a building block and exchange of two
adjacent building blocks.

Rotations are more complicated to discretise. A reasonable criterion is to
demand, on the basis of symmetry, that the allowed states be all equivalent
and equidistant from each other. The largest set of such states can then
provide an approximate basis for the rotation group, and the following
properties are quickly discovered:\\
$\bullet$ In our three-dimensional world, the largest number of equivalent
and equidistant states is four. They correspond to the corners of a regular
tetrahedron. One can go from any one to any other with equal ease---just
one step.\\
$\bullet$ A tetrahedron is the smallest polyhedron. It is the simplest
structure that can implement non-commutative features of three-dimensional
rotations. (In general, the simplest unit for tiling a $d-$dimensional space
is a simplex with $(d+1)$ vertices. It is often convenient to construct a
$d-$dimensional space as a Cartesian product of $d$ one-dimensional spaces,
but the simplex is a much more flexible unit than a hypercube.)\\
$\bullet$ To be able to specify the three-dimensional orientation
unambiguously, the building blocks should have the capability to include
a chiral center. Tetrahedral geometry allows that.\\
$\bullet$ If quantum dynamics is involved, then the states should also be
mutually orthogonal, so that they form a basis for the Hilbert space. The
tetrahedral quantum states are mutually orthogonal; they can be obtained
from the lowest two spherical harmonics, $l=0$ and $l=1$. ($l=0,1$ form
the minimal basis set for specifying orientations in three-dimensional
space. They are the smallest two representations of the group of proper
rotations, and any other representation can be constructed from them by
tensor products.) Using $sp^3-$hybridisation of atomic orbitals, these
states can be denoted as:
\begin{equation}
\left( \matrix{ \alpha\cr \beta\cr \gamma\cr \delta\cr } \right) =
\left( \matrix{ 1/2 & 1/2 & 1/2 & 1/2 \cr
		1/2 & 1/2 &-1/2 &-1/2 \cr
                1/2 &-1/2 & 1/2 &-1/2 \cr
                1/2 &-1/2 &-1/2 & 1/2 \cr } \right)
\left( \matrix{ s \cr p_x \cr p_y \cr p_z \cr } \right)
\end{equation}
The high symmetry of this unitary transformation (all elements equal, only
signs differ) is related to the equivalence of the four states.\\
$\bullet$ Four is also the largest number of states which can be uniquely
identified by a single yes/no question in a quantum search algorithm
(Grover 1997).

\section{2. Tetrahedral geometry}

The outstanding example of an element with such states is carbon. Moreover,\\
$\circ$ Carbon has the capability to form aperiodic chains, where different
side chemical groups hang on to a back-bone. This capability is a must for
encoding information. Silicon also possesses the same tetrahedral states,
and is much more abundant, but it preferentially forms periodic chains (i.e.
regular crystals).\\
$\circ$ If the logic above is repeated in the case of two-dimensional
rotations, it leads to three equivalent states located at the corners of an
equilateral triangle. Carbon has the capability to form these states as well,
by $sp^2-$hybridisation of its atomic orbitals.\\
$\circ$ Carbon is the most important structural element forming the back-bone
of biomolecules. Darwinian selection in evolution can be expected to have
picked the best building blocks out of the available resources.

With all these pieces fitting together, let us look at the tetrahedral group
in some detail. The tetrahedral group is isomorphic to the permutation group
of four objects. It has 24 elements, which can be factored into a group of
12 proper rotations (or even permutations) and reflection (or parity). The
24 element and 12 element groups are denoted as $T_d$ and $T$ respectively.

A regular tetrahedron can be formed by joining alternate corners of a cube.
The centres of the tetrahedron and cube then coincide, and this embedding is
convenient for three-dimensional structural analysis of a chain with
tetrahedral angles. The 12 proper rotations are decomposed into the identity
operation, rotations around 3-fold axes and rotations around 2-fold axes.
There are four 3-fold axes, each joining the centre of the tetrahedron with
a vertex; $+120^\circ$ and $-120^\circ$ rotations around these axes belong
to different equivalence classes. There are three 2-fold axes, each passing
through the center of the tetrahedron and midpoints of its non-intersecting
edges (equivalently passing through the centres of opposite faces of the
embedding cube).

For a carbon atom located at the centre of the tetrahedron, rotations around
3-fold axes correspond to rotations around its bonds. These single bonds are
easy to rotate and give rise to different conformations of organic molecules.
In a polypeptide chain, the orientations that can be achieved by rotations
around the bonds of the $C_\alpha$ atoms are described by the Ramachandran
map. As shown in Fig.1, the rotation angles are not uniformly populated, but
prefer to be in several discrete locations. As the stars in the plot show,
discretising the angles in steps of $120^\circ$ is not a bad starting point.

\begin{figure}[tbh]
{
\vspace*{-5mm}
\epsfxsize=10cm
\centerline{\epsfbox{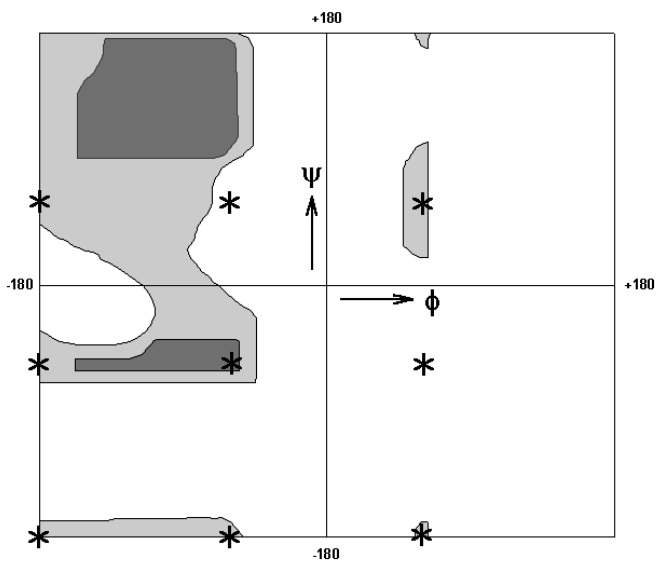}}
}
\caption{The Ramachandran map for chiral L-type amino acids, displaying the
permitted rotation angles for the $C_\alpha$ bonds in polypeptide chains
(Ramachandran 1963). The angles $\phi$ and $\psi$ are periodic.
In the approximation that embeds the polypeptide chain on a diamond lattice
in the ``trans'' configuration, only nine discrete possibilities exist for
the rotation angles. These are marked by stars on the same plot; they are
uniformly separated by $120^\circ$ steps.}
\label{fig:psiphiplot}
\end{figure}

The 2-fold rotation axes bisect the bond angles. If a double bond is viewed
as a deformation in which two tetrahedral bonds are merged together, then the
double bond lies along the 2-fold rotation axis. $180^\circ$ rotation about
this axis corresponds to a transition between ``trans'' and ``cis'' forms.
Most of the peptide bonds have the ``trans'' configuration. But occasional
transitions to the ``cis'' form do occur, and they are important for
introducing sharp bends in the chain.

The parity transformation flips chirality of a structure, which is of special
significance for many biological molecules. Chirality flip is an allowed
quantum transformation, e.g. the $NH_3$ molecule flips back and forth
between configurations where the nitrogen atom is above and below the plane
of three hydrogen atoms. But chirality flip becomes more difficult as the
molecular size increases, and all the amino acids used as building blocks
of proteins are known to be L-type (except for achiral glycine). Thus
reflections are more difficult to implement than proper rotations, and can
be ignored as far as the structural analysis of proteins is concerned.

\section{3. Packing three-dimensional information}

Multi-dimensional structural information can be encoded in several different
ways. The complete information can be expressed directly, as in holograms
(3-dim) and movie projections (2-dim). Or it can be arranged as an ordered
set of lower dimensional segments, as in CT-scan (stack of parallel planes
covering a 3-dim object) and television monitors (set of lines covering a
2-dim picture). The choice depends on whether the physical means that convey
the information are extended or local. When mechanisms exist to look at the
whole object in one go (e.g. with a wide beam of light), the complete
information can be addressed directly. When only one part of the object can
be considered at a time (e.g. with a narrow beam of electrons), it is more
convenient to arrange the information as a sequence of small segments. When
the building blocks themselves have to convey the information, the latter
format is the obvious choice; multi-dimensional arrays are stored as folded
sequences in computers and proteins are assembled as folded polypeptide chains. 

It is possible to assemble arbitrary structures by repetitive arrangement
of a single and small enough building block. For example, a crystal can be
carved into the desired shape, and it is sufficient to describe the details
of the surface (and not the contents of the full volume) for that purpose.
A crystal can also form rapidly, since it can grow from a seed in all
directions. But the preferred shape of the crystal remains that of the
building block. To assemble arbitrary shapes using a crystalline arrangement,
another agency is needed to tell the crystal surface to stop growing after
it has reached the desired position as well as to put the reactive chemical
groups at specific locations; the building blocks themselves cannot carry
those instructions. Thus crystal growth is convenient for making regular
patterns, but it is not a good choice for assembling irregular shapes.

The highly non-trivial task of specifying an irregular structure can be more
easily accomplished by an aperiodic folded chain of building blocks. Then
the building blocks themselves carry preferences for specific orientations
at each step. Although such a chain grows slowly, it does not need help from
an external agency to achieve its desired shape. This property is a must at
the lowest level of information processing---the message has to carry its
own interpretation in terms of its physical properties; no other interpreter
is available\footnote{%
In case of computers, compilers and operating systems provide the abstract
interpretation for high level information processing. But at the lowest level
of machine code, the interpretation is built into the design of the physical
components, i.e. in their responses to applied voltages and currents.}.
With a chain that knows how to fold itself, the problem of specifying the
three-dimensional structural detail is simplified to that of constructing
the appropriate one-dimensional chain. It is far easier for an external
agency to synthesise aperiodic one-dimensional chains than irregular
three-dimensional structures. Proteins do not have regular shapes; they
need all their grooves and cavities (i.e. structural defects) for their
function, and how they fold is decided by their building blocks joined in
a polypeptide chain. Another physical reason why proteins have to be
polypeptide chains that can fold and unfold again is that many proteins
have to cross membranes and cell walls to carry out their tasks. A bulky
shape would require a big hole in the barrier to be crossed, through which
many other molecules could also leak. But proteins unfold to their chain
form, slip through a small hole in the barrier, and then fold again to
their native form.

Even after picking a folded chain structure, more specifications are needed
to find the desired building blocks. The chain can be uniformly flexible
like a piece of string, or it can be made of stiff segments alternating with
flexible joints like a chain of metal rings. If all the segments of a chain
are flexible, then it has to be fully tied from all directions to be held in
place. Otherwise the structure can crumple and collapse. Carbon forms many
structures with fully saturated bonds, but a completely tied three-dimensional
form requires rather precise folding and cannot accommodate aperiodic building
blocks easily. Moreover, for a chain to have the capacity to fold and unfold
again, the side bonds holding the folds in the chain must be weaker than the
bonds along the back-bone. For example, diamond is the hardest material, but
it is a periodic structure and cannot be folded and unfolded again easily.
The polyethylene back-bone (i.e. $(-CH_2-)_n$) can accommodate aperiodic side
groups, but it is too flexible. Given that the side group interactions are
necessarily weak, structural stability can be enhanced by making the back-bone
stiffer, e.g. by replacing some of the single bonds with double bonds that
cannot rotate. Polypeptide chains are of this type; they have weak side group
interactions, and the increased stiffness of their non-rotatable peptide bonds
helps in maintaining the shape of the protein.

Fig.2a shows a back-bone with alternating single and double bonds. This is
the structure of polyacetylene, and an aperiodic chain can be constructed by
replacing the side $-H$ by other chemical groups (e.g. $-CH_3$). The trouble
with this structure is that the $\pi-$electrons involved in double bonds
prefer to lower their energy by spilling over into neighbouring bonds. This
resonance phenomenon gives a double bond character to all the bonds (the
actual bond properties are somewhere in between a single and a double bond),
and makes the whole back-bone planar. A planar back-bone is no good for
constructing three-dimensional structures. The double bonds can be shifted to
the side groups to reduce the spill over of $\pi-$ electrons, as illustrated
in Fig.2b, but the resultant structure is still planar. The next possibility
for the back-bone configuration, which allows stiff segments with flexible
joints, is to alternate one double bond with two single bonds. This is the
structure of polypeptide chains, as shown in Fig.2c. The stiff $C-N$ peptide
bond is created by $\pi-$electrons spilling over from the $C=O$ double bond;
inclusion of nitrogen atoms in the chain ensures that the $\pi-$electrons
spill over only on one side and not the other. The rotatable single bonds
of $C_\alpha$ atoms permit construction of three-dimensional structures.

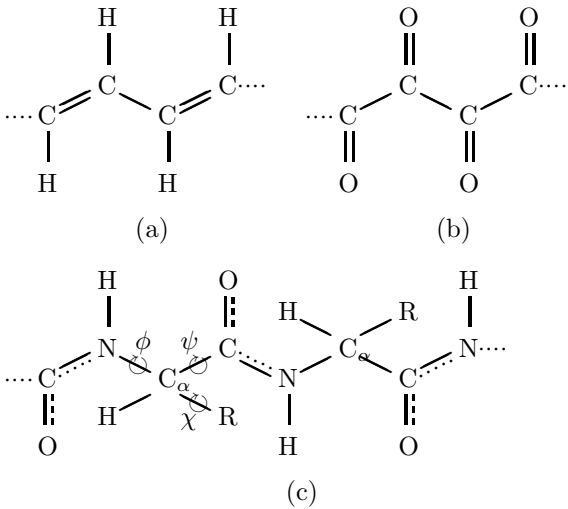
\begin{figure}[tbh]
{
\setlength{\unitlength}{1mm}
\begin{picture}(80,70)
  \thicklines
  \put(20,38){\makebox(0,0)[bl]{(a)}}
  \put( 3,54){\circle*{0.5}} \put( 4,54){\circle*{0.5}}
  \put( 5,54){\circle*{0.5}} \put( 6,54){\circle*{0.5}}
  \put( 7,53){\makebox(0,0)[bl]{C}}
  \put( 8.5,48){\line( 0,1){4}}
  \put( 7,44){\makebox(0,0)[bl]{H}}
  \put(10,54.5){\line( 2,1){4}}
  \put(10,55.5){\line( 2,1){4}}
  \put(15,57){\makebox(0,0)[bl]{C}}
  \put(16.5,61){\line( 0,1){4}}
  \put(15,66){\makebox(0,0)[bl]{H}}
  \put(22,55){\line(-2,1){4}}
  \put(23,53){\makebox(0,0)[bl]{C}}
  \put(24.5,48){\line( 0,1){4}}
  \put(23,44){\makebox(0,0)[bl]{H}}
  \put(26,54.5){\line( 2,1){4}}
  \put(26,55.5){\line( 2,1){4}}
  \put(31,57){\makebox(0,0)[bl]{C}}
  \put(32.5,61){\line( 0,1){4}}
  \put(31,66){\makebox(0,0)[bl]{H}}
  \put(34,58){\circle*{0.5}} \put(35,58){\circle*{0.5}}
  \put(36,58){\circle*{0.5}} \put(37,58){\circle*{0.5}}

  \put(60,38){\makebox(0,0)[bl]{(b)}}
  \put(43,54){\circle*{0.5}} \put(44,54){\circle*{0.5}}
  \put(45,54){\circle*{0.5}} \put(46,54){\circle*{0.5}}
  \put(47,53){\makebox(0,0)[bl]{C}}
  \put(48,48){\line( 0,1){4}}
  \put(49,48){\line( 0,1){4}}
  \put(47,44){\makebox(0,0)[bl]{O}}
  \put(50,55){\line( 2,1){4}}
  \put(55,57){\makebox(0,0)[bl]{C}}
  \put(56,61){\line( 0,1){4}}
  \put(57,61){\line( 0,1){4}}
  \put(55,66){\makebox(0,0)[bl]{O}}
  \put(62,55){\line(-2,1){4}}
  \put(63,53){\makebox(0,0)[bl]{C}}
  \put(64,48){\line( 0,1){4}}
  \put(65,48){\line( 0,1){4}}
  \put(63,44){\makebox(0,0)[bl]{O}}
  \put(66,55){\line( 2,1){4}}
  \put(71,57){\makebox(0,0)[bl]{C}}
  \put(72,61){\line( 0,1){4}}
  \put(73,61){\line( 0,1){4}}
  \put(71,66){\makebox(0,0)[bl]{O}}
  \put(74,58){\circle*{0.5}} \put(75,58){\circle*{0.5}}
  \put(76,58){\circle*{0.5}} \put(77,58){\circle*{0.5}}

  \put(40,3){\makebox(0,0)[bl]{(c)}}
  \put( 3,19){\circle*{0.5}} \put( 4,19){\circle*{0.5}}
  \put( 5,19){\circle*{0.5}} \put( 6,19){\circle*{0.5}}
  \put( 7,18){\makebox(0,0)[bl]{C}}
  \put( 8,13){\line( 0,1){4}}
  \put( 9,13){\line( 0,1){0.8}}
  \put( 9,14.6){\line( 0,1){0.8}} \put( 9,16.2){\line( 0,1){0.8}}
  \put( 7, 9){\makebox(0,0)[bl]{O}}
  \put(10,20.5){\line( 2,1){4}}
  \put(10,19.5){\circle*{0.5}} \put(11,20){\circle*{0.5}}
  \put(12,20.5){\circle*{0.5}} \put(13,21){\circle*{0.5}}
  \put(14,21.5){\circle*{0.5}}
  \put(15,22){\makebox(0,0)[bl]{N}}
  \put(16.5,26){\line( 0,1){4}}
  \put(15,31){\makebox(0,0)[bl]{H}}
  \put(22,20){\line(-2,1){4}}
  \put(19,20){$\circlearrowright$}
  \put(20,23){$\phi$}
  \put(23,18){\makebox(0,0)[bl]{C$_\alpha$}}
  \put(22,17){\line(-2,-1){4}}
  \put(26,17){\line( 2,-1){4}}
  \put(15,13){\makebox(0,0)[bl]{H}}
  \put(27,15){$\circlearrowright$}
  \put(26,13){$\chi$}
  \put(31,13){\makebox(0,0)[bl]{R}}
  \put(26,20){\line( 2,1){4}}
  \put(27,20){$\circlearrowright$}
  \put(26,23){$\psi$}
  \put(31,22){\makebox(0,0)[bl]{C}}
  \put(32,26){\line( 0,1){4}}
  \put(33,26){\line( 0,1){0.8}}
  \put(33,27.6){\line( 0,1){0.8}} \put(33,29.2){\line( 0,1){0.8}}
  \put(31,31){\makebox(0,0)[bl]{O}}
  \put(38,19.5){\line(-2,1){4}}
  \put(38,20.5){\circle*{0.5}} \put(37,21){\circle*{0.5}}
  \put(36,21.5){\circle*{0.5}} \put(35,22){\circle*{0.5}}
  \put(34,22.5){\circle*{0.5}}
  \put(39,18){\makebox(0,0)[bl]{N}}
  \put(40.5,13){\line( 0,1){4}}
  \put(39, 9){\makebox(0,0)[bl]{H}}
  \put(42,20){\line( 2,1){4}}
  \put(47,22){\makebox(0,0)[bl]{C$_\alpha$}}
  \put(46,25){\line(-2, 1){4}}
  \put(50,25){\line( 2, 1){4}}
  \put(39,27){\makebox(0,0)[bl]{H}}
  \put(55,27){\makebox(0,0)[bl]{R}}
  \put(54,20){\line(-2,1){4}}
  \put(55,18){\makebox(0,0)[bl]{C}}
  \put(56,13){\line( 0,1){4}}
  \put(57,13){\line( 0,1){0.8}}
  \put(57,14.6){\line( 0,1){0.8}} \put(57,16.2){\line( 0,1){0.8}}
  \put(55, 9){\makebox(0,0)[bl]{O}}
  \put(58,20.5){\line( 2,1){4}}
  \put(58,19.5){\circle*{0.5}} \put(59,20){\circle*{0.5}}
  \put(60,20.5){\circle*{0.5}} \put(61,21){\circle*{0.5}}
  \put(62,21.5){\circle*{0.5}}
  \put(63,22){\makebox(0,0)[bl]{N}}
  \put(64.5,26){\line( 0,1){4}}
  \put(63,31){\makebox(0,0)[bl]{H}}
  \put(66,23){\circle*{0.5}} \put(67,23){\circle*{0.5}}
  \put(68,23){\circle*{0.5}} \put(69,23){\circle*{0.5}}
\end{picture}
}
\caption{Different possibilities for polymer chains with carbon back-bone:
(a) $(CH)_n$, (b) $(CO)_n$, (c) polypeptide chain.}
\label{fig:Cchains}
\end{figure}

\begin{figure}[tbh]
{
\setlength{\unitlength}{1mm}
\begin{picture}(80,32)
  \thicklines
  \put(10,2){\makebox(0,0)[bl]{(a)}}
  \put(11,19){\makebox(0,0)[bl]{C}}
  \put(10,20){\line(-1, 0){4}}
  \put(-2,19){\makebox(0,0)[bl]{H$_3^+$N}}
  \put(14,20){\line( 1, 0){4}}
  \put(19,19){\makebox(0,0)[bl]{H}}
  \put(12,22){\line( 0, 1){4}}
  \put(11,27){\makebox(0,0)[bl]{COO$^-$}}
  \put(12,18){\line( 0,-1){4}}
  \put(11,11){\makebox(0,0)[bl]{H}}

  \put(39,2){\makebox(0,0)[bl]{(b)}}
  \put(40,21.5){\makebox(0,0)[bl]{C}}
  \put(43,24){\line( 2, 1){4}}
  \put(47.5,27){\makebox(0,0)[bl]{H}}
  \put(40,24){\line(-2, 1){4}}
  \put(33,27){\makebox(0,0)[bl]{COO$^-$}}
  \put(43,21){\line( 2,-1){4}}
  \put(47.5,16){\makebox(0,0)[bl]{CH$_2$}}
  \put(40,21){\line(-2,-1){4}}
  \put(28,16){\makebox(0,0)[bl]{H$_2^+$N}}
  \put(49,15){\line(0,-1){4}}
  \put(47.5,8){\makebox(0,0)[bl]{CH$_2$}}
  \put(34,15){\line(0,-1){4}}
  \put(28.5,8){\makebox(0,0)[bl]{H$_2$C}}
  \put(36,9){\line(1,0){10}}

  \put(71,2){\makebox(0,0)[bl]{(c)}}
  \put(72,19){\makebox(0,0)[bl]{C}}
  \put(71,20){\line(-1, 0){4}}
  \put(59,19){\makebox(0,0)[bl]{H$_3^+$N}}
  \put(75,20){\line( 1, 0){4}}
  \put(80,19){\makebox(0,0)[bl]{H}}
  \put(73,22){\line( 0, 1){4}}
  \put(72,27){\makebox(0,0)[bl]{COO$^-$}}
  \put(73,18){\line( 0,-1){4}}
  \put(72,11){\makebox(0,0)[bl]{R}}
\end{picture}
}
\caption{Amino acid configurations (in their ionised forms):
(a) glycine, (b) proline, (c) all the rest.}
\label{fig:Rgroups}
\end{figure}
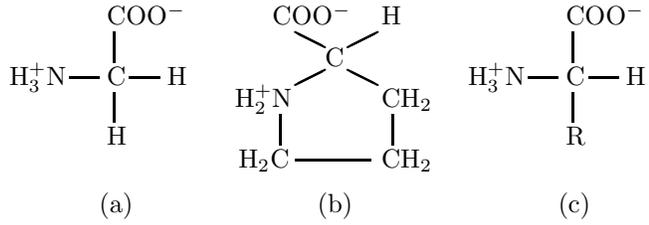

The configurations of amino acids that polymerise to form the polypeptide
chain are shown in Fig.3. The presence of acidic $-COOH$ and basic $-NH_2$
groups in all amino acids provides a convenient way to join them in
polypeptide chains by acid-base neutralisation.

\section{4. Elementary building blocks}

Having analysed the merits of a polypeptide back-bone structure,
we now look at the three-dimensional geometry of a polypeptide chain,
but with the simplifying assumptions that all the links in the chain
are of equal length and all the tetrahedral angles are of equal value
($2\tan^{-1}(\sqrt{2}) \approx 109.5^\circ$). With these assumptions,
the folded chain lies on a diamond lattice. Although the real peptide bond
is planar with angles close to $120^\circ$, it can be fitted reasonably
well on the diamond lattice in the ``trans'' configuration (see Fig.4).
The rare ``cis'' configuration, takes the chain out of the diamond lattice.
Let us first keep the ``cis'' configuration aside, and consider the chain
in the ``trans'' configuration only. In a real polypeptide chain, variations
from equal bond lengths and equal angles are within $\pm10\%$, and I will
analyse the above described simplified version using the conventional
polypeptide chain nomenclature.

The diamond lattice is a face-centred cubic lattice with a two-point basis.
Let this basis of lattice points be $(0,0,0)$ and $(1/4,1/4,1/4)$ in units
of the unit cell. Then the bond directions of the diamond lattice are (these
are the last three columns of the matrix in Eq.(1)):
\begin{eqnarray}
e_1 &= (+1/4,+1/4,+1/4) \cr
e_2 &= (+1/4,-1/4,-1/4) \cr
e_3 &= (-1/4,+1/4,-1/4) \cr
e_4 &= (-1/4,-1/4,+1/4)
\end{eqnarray}
These directions refer to the lattice point at the origin, and thereafter
the bond directions at neighbouring points are opposite in sign.

We can enumerate all possible configurations of the polypeptide chain, by
specifying for every peptide bond the location of the next peptide bond in
the chain. Let the reference peptide bond ($C-N$) be along $e_1$ from the
origin. The chain prior to this peptide bond is already synthesised, so
without loss of generality let the location of the $C_\alpha$ preceding
the reference peptide bond be $e_2$. In the ``trans'' configuration, the
$N-C_\alpha$ and the $C_\alpha-C$ bonds are parallel, so the location of
the $C_\alpha$ following the reference peptide bond is fixed as $e_1-e_2$.
(The sequence $C_\alpha-C-N-C_\alpha$ fixes the plane of the peptide bond.)

There are three possible locations for the next $C$: $e_1-e_2+e_1$,
$e_1-e_2+e_3$ and $e_1-e_2+e_4$. From each of these three locations, the
next peptide bond can proceed along three possible directions, excluding the
already occupied $C_\alpha-C$ direction. The $C_\alpha-C$ direction and the
next peptide bond direction fix the plane of the next peptide bond. Thus
on a diamond lattice, given a peptide bond plane, there are $9$ possible
positions for the next peptide bond plane.

These $9$ orientations are all geometrically equivalent---they just reflect
the 3-fold rotational symmetry around tetrahedral bonds of the $C_\alpha$
atom. When $C_\alpha-N$ bond is held in position, there are three equivalent
choices of $\phi$, and when $C_\alpha-C$ bond is held in position, there
are three equivalent choices of $\psi$.

For each of the polypeptide back-bone configuration described above, there are
two remaining directions for other groups to attach to the $C_\alpha$ atom.
One direction is attached to the R-group of the amino acid, while the other
to a hydrogen atom. There are two arrangements possible, and they correspond
to opposite chirality. Detailed model-building studies have shown that all
the R-groups in a polypeptide chain must be of the same stereoisomer for the
stability of regular secondary structures (e.g. $\alpha-$helices and
$\beta-$sheets). All the amino acids naturally occurring in proteins are L-type.
Altogether, therefore, there remain $9$ possible ways of adding a new L-type
amino acid to an existing polypeptide chain. (Instead of an R-group, glycine has
two hydrogen atoms attached to the $C_\alpha$ atom. That makes glycine achiral,
but the number of attachment possibilities for the $C_\alpha$ atom remains one.)

The Ramachandran map shown in Fig.1 is constructed using the bond lengths
and angles in an actual polypeptide chain. The nine points corresponding to
the ``trans'' configuration discrete chain are marked as stars on the same
plot. It is easily seen that the discrete approximation is not too far off
reality, even though it cannot describe all the details of the $\phi-\psi$
angular distribution. Actually, the plot in Fig.1 does not include glycine
and proline; their structural preferences are somewhat different. The region
around $(\phi=60^\circ,\psi=-60^\circ)$ is not occupied in the Ramachandran
map because of steric conflict between the side chain R-group and the atoms
in the polypeptide back-bone. Glycine with no side chain does not have this
conflict and can occupy this region---its Ramachandran map has inversion
symmetry. In case of proline, the rigid imino ring does not allow the
$N-C_\alpha$ bond to rotate, and $\phi$ is constrained to be around
$-60^\circ$. In case of a real polypeptide chain, embedding it on the
diamond lattice will distort its shape; the extent of distortion will then
be a measure of usefulness of the discretised description. (The peptide bond
is a little shorter than the single bonds and its bond angles of $120^\circ$
are somewhat wider than the tetrahedral angle. These two deviations tend to
compensate for each other to some extent.)

Now we can look at the ``cis'' configuration of the peptide bond. It is
obtained from the ``trans'' configuration by rotating the $N-C_\alpha$
bond by $180^\circ$ around the peptide bond axis (see Fig.4). With the
peptide bond along $e_1$ and the preceding $C_\alpha-C$ bond along $-e_2$,
the ``cis'' configuration $N-C_\alpha$ bond is along ${\small 2 \over 3}
e_1 + e_2$. This orientation does not fit in the face-centred cubic diamond
lattice, but it can be fitted in the hexagonal diamond lattice\footnote{%
Carbon can also form a hexagonal diamond lattice, with the same tetravalent
bonds and density as the face-centred cubic diamond lattice. Such hexagonal
diamond crystals do not occur terrestrially, but they have been found in
meteorites and have been synthesised in laboratory.}
with the hexagonal symmetry axis along $e_1$. It is well-known that the
three-dimensional closest packing of spheres can be viewed as a stack of
two-dimensional layers. There are three possible positions for the layers,
and each layer has to be displaced relative to the ones on its either side.
There are, therefore, two distinct ways to add a new layer onto an existing
stack. The face-centred cubic lattice corresponds to the layer sequence
$\ldots ABCABCABC \ldots$, the hexagonal lattice corresponds to the sequence
$\ldots ABABAB \ldots$, and random sequences are also possible. An insertion
of a ``cis'' peptide bond in an otherwise ``trans'' peptide chain corresponds
to a flip in the layer sequence of the type $\ldots ABCABCBACBA \ldots$.
This flip has no effect on the $9$ possibilities for the subsequent rotation
angles $\phi$ and $\psi$, and further elongation of the polypeptide chain.
Thus we can count the trans-cis transformation as one more elementary
structural operation.

The 10 operations described above exhaust the ``elementary logic gates'' for
the polypeptide chain embedded on a diamond lattice, i.e. by implementing
these 10 operations one can fold the polypeptide chain on a diamond lattice
in any desired configuration. Long distance connections (disulfide and
hydrogen bonds) are important for structural stability of the polypeptide
chain, but they do not give rise to new configurational possibilities.

\begin{figure}[tbh]
{
\setlength{\unitlength}{1mm}
\begin{picture}(80,30)
  \thicklines
  \put(20,2){\makebox(0,0)[bl]{(a)}}
  \put( 4,16){\circle*{0.5}} \put( 5,16){\circle*{0.5}}
  \put( 6,16){\circle*{0.5}} \put( 7,16){\circle*{0.5}}
  \put( 8,15){\makebox(0,0)[bl]{C$_\alpha$}}
  \put(11,17){\line( 2,1){4}}
  \put(16,19){\makebox(0,0)[bl]{C}}
  \put(17,23){\line( 0,1){4}}
  \put(18,23){\line( 0,1){0.8}}
  \put(18,24.6){\line( 0,1){0.8}} \put(18,26.2){\line( 0,1){0.8}}
  \put(16,28){\makebox(0,0)[bl]{O}}
  \put(23,16.5){\line(-2,1){4}}
  \put(23,17.5){\circle*{0.5}} \put(22,18){\circle*{0.5}}
  \put(21,18.5){\circle*{0.5}} \put(20,19){\circle*{0.5}}
  \put(19,19.5){\circle*{0.5}}
  \put(24,15){\makebox(0,0)[bl]{N}}
  \put(25.5,10){\line( 0,1){4}}
  \put(24, 6){\makebox(0,0)[bl]{H}}
  \put(27,17){\line( 2,1){4}}
  \put(32,19){\makebox(0,0)[bl]{C$_\alpha$}}
  \put(37,20){\circle*{0.5}} \put(38,20){\circle*{0.5}}
  \put(39,20){\circle*{0.5}} \put(40,20){\circle*{0.5}}

  \put(60,2){\makebox(0,0)[bl]{(b)}}
  \put(44,16){\circle*{0.5}} \put(45,16){\circle*{0.5}}
  \put(46,16){\circle*{0.5}} \put(47,16){\circle*{0.5}}
  \put(48,15){\makebox(0,0)[bl]{C$_\alpha$}}
  \put(51,17){\line( 2,1){4}}
  \put(56,19){\makebox(0,0)[bl]{C}}
  \put(57,23){\line( 0,1){4}}
  \put(58,23){\line( 0,1){0.8}}
  \put(58,24.6){\line( 0,1){0.8}} \put(58,26.2){\line( 0,1){0.8}}
  \put(56,28){\makebox(0,0)[bl]{O}}
  \put(63,16.5){\line(-2,1){4}}
  \put(63,17.5){\circle*{0.5}} \put(62,18){\circle*{0.5}}
  \put(61,18.5){\circle*{0.5}} \put(60,19){\circle*{0.5}}
  \put(59,19.5){\circle*{0.5}}
  \put(64,15){\makebox(0,0)[bl]{N}}
  \put(67,17){\line( 2,1){4}}
  \put(72,19){\makebox(0,0)[bl]{H}}
  \put(65.5,10){\line( 0,1){4}}
  \put(64, 6){\makebox(0,0)[bl]{C$_\alpha$}}
  \put(69, 7){\circle*{0.5}} \put(70, 7){\circle*{0.5}}
  \put(71, 7){\circle*{0.5}} \put(72, 7){\circle*{0.5}}
\end{picture}
}
\caption{Peptide bond configurations: (a) trans, (b) cis.}
\label{fig:Pbond}
\end{figure}
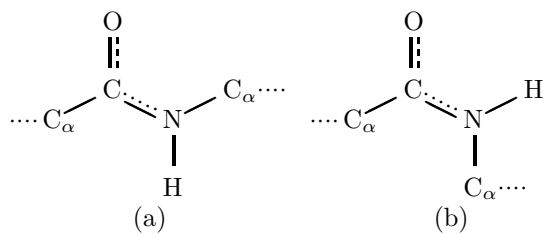

\newpage
\section{5. Putting things together}

There is no clear association of any amino acid with the 9 discrete points
in the Ramachandran map. Even the most rigid proline occurs in different
orientations, and though only glycine can occupy the region around
$(\phi=60^\circ,\psi=-60^\circ)$ it occurs in other orientations too.
Indeed, just the composition of a particular amino acid does not decide its
configuration in the polypeptide chain; rather the overall interactions of
its R-group with those that precede it and those that follow it fix the
configuration. When the orientation of the middle amino acid depends on the
amino acids that precede and follow it, the structural code is necessarily
an overlapping one. Even with an overlapping structural code, every time an
amino acid is added to the polypeptide chain, the orientation of one amino
acid gets decided. So a maximally overlapping efficient code needs at least
9 amino acids (may be 10 to include trans-cis transformation) to construct
a polypeptide chain of arbitrary configuration\footnote{%
Information is localised in individual building blocks in a strictly local
code (e.g. our system of writing numbers), but it is not so in an overlapping
code (e.g. pronunciation of vowels in English words often depends on the
neighbouring letters). As long as the total information content of the message
remains the same, one can map the two codes into each other by changing
variables. When both the codes are efficient, the total length of the message
and the total information content remain the same, and the number of building
blocks required cannot change. One can reduce the number of building blocks at
the expense of increasing the length of the message, but that would not make
the best use of available resources.}.
The R-group properties of amino acids have been studied in detail: polar
and non-polar, positive and negative charge, straight chains and rings,
short and long chains, and so on. Still which sequence of amino acids will
lead to which conformation of the chain is an exercise in coding that has
not been solved yet.

\begin{table}[tbh]
\begin{tabular}{|l|l|r|r|}
\hline
Amino acid          & R-group property & Mol. wt. & Class \\
\hline
Gly (Glycine)       & Non-polar        & 75       & II \\  
Ala (Alanine)       & Non-polar        & 89       & II \\  
Pro (Proline)       & Non-polar        & 115      & II \\  
Val (Valine)        & Non-polar        & 117      & I  \\  
Leu (Leucine)       & Non-polar        & 131      & I  \\  
Ile (Isoleucine)    & Non-polar        & 131      & I  \\  
\hline
Ser (Serine)        & Polar            & 105      & II \\  
Thr (Threonine)     & Polar            & 119      & II \\  
Asn (Asparagine)    & Polar            & 132      & II \\  
Cys (Cysteine)      & Polar            & 121      & I  \\  
Met (Methionine)    & Polar            & 149      & I  \\  
Gln (Glutamine)     & Polar            & 146      & I  \\  
\hline
Asp (Aspartate)     & Negative charge  & 133      & II \\  
Glu (Glutamate)     & Negative charge  & 147      & I  \\  
\hline
Lys (Lysine)        & Positive charge  & 146      & II \\  
Arg (Arginine)      & Positive charge  & 174      & I  \\  
\hline
His (Histidine)     & Ring/Aromatic    & 155      & II \\  
Phe (Phenylalanine) & Ring/Aromatic    & 165      & II \\  
Tyr (Tyrosine)      & Ring/Aromatic    & 181      & I  \\  
Trp (Tryptophan)    & Ring/Aromatic    & 204      & I  \\  
\hline
\end{tabular}
\caption{Properties of the amino acids depend on their side chain R-groups.
Larger molecular weights indicate longer side chains. The 20 amino acids
naturally occurring in proteins have been divided into two classes of 10 each,
depending on the properties of aminoacyl-tRNA synthetases that bind the amino
acids to tRNA. These classes divide amino acids with each R-group property
equally, the longer side chains correspond to class I and the shorter ones
correspond to class II. Some specific properties not explicit in the table
are: asparagine is a shorter side chain version of glutamine, histidine has
an R-group with a small positive charge but it is close to being neutral,
and both the sulphur containing amino acids (cysteine and methionine) belong
to class I.}
\end{table}

At this stage, it is instructive to observe that the 20 amino acids are
divided into two classes of 10 each, according to the properties of their
aminoacyl-tRNA synthetases
(Eriani {\it et al.} 1990, Arnez and Moras 1997, Lewin 2000).
The two classes of synthetases totally differ from each other in their
active sites and in how they attach amino acids to the tRNA molecules. The
lack of any apparent relationship between the two classes of synthetases
has led to the conjecture that the two classes evolved independently, and
early form of life could have existed with proteins made up of only 10
amino acids of one type or the other. A closer inspection of the R-group
properties of amino acids in the two classes reveals that each property
(polar, non-polar, ring/aromatic, positive and negative charge) is equally
divided amongst the two classes, as shown in Table 1. Not only that, but the
heavier amino acids with each property belong to class I, while the lighter
ones belong to class II. This division of amino acids according to the length
of their side chains has unambiguous structural significance. The diamond
lattice structure is quite loosely packed with many cavities of different
sizes. The use of long side chains to fill up big cavities and short side
chains to fill up small ones can produce a dense compact structure, and
proteins indeed are tightly packed close to the maximum packing fraction.
Thus we arrive at a structural explanation for the 20 amino acids as building
blocks of proteins, a factor of 10 for conformations of the polypeptide
back-bone and a factor of 2 for the length of the R-group. (It can be noted
that each class contains one special amino acid involved in tasks beyond the
9 folding possibilities on the diamond lattice. Proline is the special amino
acid in class II, involved in trans-cis transformations, while cysteine is
the special amino acid in class I, involved in tying together far separated
regions of the polypeptide chain.)

A look at the optimal solutions of the quantum search algorithm (Grover 1997)
brings out another interesting feature. Living organisms form DNA and
polypeptide chains by joining their building blocks together in sequential
assembly line operations. The templates for the assembly, hereditary DNA
and mRNA, are preexisting objects in these processes. The building blocks
are available as a random collection, and the correct ones are selected by
ensuring that they form appropriate molecular bonds with the templates.
Molecular bonds are binary questions---they either form or do not form.
The optimisation criterion is to select the correct building block from
the random collection by asking the minimum number of questions. With binary
questions, the best classical search algorithm is a binary search, but the
best quantum search algorithm is different. Identification of the nucleotide
base-pairing with a binary quantum question provided two significant results
for genetic information processing (Patel 2000):
the largest number of items that can be distinguished by one quantum question
is 4, and by three quantum questions is 20.2. These numbers match the number
of letters in DNA and protein alphabets, and the triplet code translating
between them. The same algorithm also predicts that the largest number of items
that can be distinguished by two quantum questions is 10.5. (The non-integer
number of items means that the algorithm has an intrinsic error. In the case
of two queries and 10 items, the error rate is about 1 part in 1000.)
A two nucleotide base code is thus optimal for distinguishing 10 amino acids.

The experimentally observed wobble rules are consistent with the idea that
an earlier genetic code used only two nucleotide bases of every codon and
synthesised a smaller number of amino acids (Crick 1966).
Another feature supporting this idea is that in the present genetic code
similar codons code for amino acids with similar R-group properties. It is
therefore possible that the third codon entered the present genetic code
as a class label (classical and not quantum), when two independent codes
corresponding to long and short R-groups merged together during the course
of evolution. Such symbiosis would not be uncommon---there is evidence that
the cellular organelles mitochondria and chloroplasts, with their own genetic
material, first developed independently and were later incorporated in
ancestral cells with eukaryotic nuclei.

Combining all these arguments, we can now construct a possible scenario of
how the present genetic code arose from a more primitive one. Since the
discovery of the genetic code, many attempts have been made to find its
simpler predecessors (see for instance, Crick 1968, Kolaskar and Ramabrahmam
1982, Ikehara 2002). The present genetic code is too complex to have arisen
in one go, and all the simpler predecessors use fewer nucleotide bases and
fewer amino acids. The most important criterion in these attempts is that
continuity has to be maintained in evolution---a drastic change will not
permit the organism to survive. The changes therefore have to combine many
small steps, each small change in the code providing a certain advantage
in functionality. The scenario suggested by the preceding arguments is
similar to what Crick proposed many years ago (Crick 1968):\\
(1) The primitive code was a triplet one due to some unidentified reasons.
The first two letters coded for 10 amino acids, while the third letter was
a non-coding separation mark. The individual genes were separate and not
joined together, and so START and STOP signals were not needed.\\
(2) This primitive code synthesised the simpler class II amino acids.
The information about how the polypeptide chain twists and turns at each
step was incorporated in this code. The short side chains of class II
amino acids, however, could not completely fill all the cavities in the
three-dimensional protein structure.\\
(3) The longer class I amino acids replaced the short ones of similar property
at a later stage, wherever big cavities existed. This filling up of cavities
increased the structural stability of proteins.\\
(4) The third letter was put into use as a double-valued classical label for
the amino acid class. That allowed coding for 20 amino acids.\\
(5) Further optimisation of the code occurred with some juggling of codons,
since 20 amino acids can be coded either by one classical and two quantum
queries or by three quantum queries. Also, many genes joined together and
START and STOP signals were inserted.\\
(6) Similar codons for similar amino acids and the wobble rules are relics
of the doubling of the genetic code, indicative of the past but no longer
perfectly realised.\\
In the absence of any knowledge of the doublet code, it is not possible to
pin-point this scenario any further, and variations can be imagined.

Further progress along this direction requires solutions of two puzzles.
First, as already pointed out above, we need to identify which amino acid
subsequence corresponds to which structural building block. There is no
clear criterion regarding how long the amino acid subsequence should be
before it assumes a definite shape; may be interactions of an amino acid
with two preceding ones and two following ones is a good enough beginning.
The protein structure data accumulated in databases should help in such an
analysis. Second, we need to guess the doublet code assignments from the
known triplet code. This has already been studied to some extent within
the context of the wobble rules, but it should be investigated in more
detail keeping the constraints of the aminoacyl-tRNA synthetase classes
in mind.

\section{6. Summary and outlook}

I have looked at the structure of proteins from an information theory point
of view. The emphasis is on the three-dimensional structure of the end-product,
i.e. how should the segments of a polypeptide chain be chosen so that it folds
into the required shape. The means used to achieve that end are secondary,
i.e. which amino acids should be chosen so that the interactions amongst
their R-groups make the polypeptide chain fold in the required manner. This
emphasis is in sharp contrast with the conventional approach to the protein
folding problem, i.e. find the three-dimensional structure of the protein,
given the sequence of amino acids and the interactions of their R-groups.
The conventional approach requires finding the lowest energy configuration
of a polypeptide chain, and is believed to be NP-hard, because finding the
global energy minimum with all possible interactions is not at all easy.
The rephrased problem of structural design may not be that hard---the
local orientation of a building block can be fixed by its interactions
with its neighbours; it is enough to have a locally stable or metastable
configuration and not necessarily a global energy minimum. (Diamond is
structurally the strongest material, but it is energetically metastable.)
Also, there are many ways a folded chain can cover a three-dimensional shape,
and quite likely there is a lot of flexibility in choosing the sequence of
amino acids without substantially altering the structure of the protein.

The fundamental operations needed for processing structural information
are translation and rotation. I have shown that carbon and its tetrahedral
geometry provide the simplest discretisation of these operations. For the
construction of proteins as folded chains, the polypeptide chain is the
simplest back-bone containing rigid segments alternating with flexible joints.
To fold this back-bone into arbitrary shapes on a diamond lattice requires
10 basic operations. The amino acids somehow implement these operations by
interactions amongst their side chain R-groups.

I have pointed out that the division of the 20 amino acids, by aminoacyl-tRNA
synthetases, into two classes of 10 each has structural significance. Every
R-group property is equally divided between the two classes, such that the
shorter side chains are in class II and the longer ones in class I. This is
a new observation. Combining this fact with the number of discrete operations
required to fold a polypeptide chain, and the result that two yes/no quantum
queries can distinguish 10 items, I have proposed that the present triplet
genetic code was preceded by a primitive doublet one. How the doublet code
was converted to a triplet one is a matter of conjecture, and I have outlined
one possible scenario.

Knowing the solution selected by evolution has no doubt guided my logic.
Still unraveling the optimisation criteria involved in the design of
molecules of life is a thrilling exercise. It should be kept in mind that
evolution has discovered its optimal parameters, not by logical deduction,
but by trial and error experiments (of course using the available means).
For that reason the chosen parameters are not always perfect. On the other
hand, evolution has had plenty of time for experimentation, something which
we do not, and cannot, have. As a result, though evolution is not perfect
in finding its criteria, it is impressive to say the least!

A shortcoming of my analysis is that the role played by water in protein
folding is totally ignored. Water molecules do not provide just a uniform
background inside the cells; they fit in tetrahedral geometry nicely, and
are therefore well-suited to fill up empty grooves and cavities of proteins.
Any explanation of interactions amongst the side chain R-groups of amino
acids must include how the R-groups interact with water. This unfinished
exercise is for the future.

The ideas discussed in this work have potential applications. Molecular
dynamics simulations of protein folding have typically been carried out in
the continuous three-dimensional space. The discrete information theory
language can speed up these simulations, by first folding the polypeptide
chain on a diamond lattice and then switching to the continuous space to
fine tune the actual atomic positions. In recent years, lattice models
have been used for the analysis of polypeptide folds. But they have mostly
used cubic or triangular lattices and restricted amino acid properties to
a binary hydrophobic/hydrophilic choice (Hart and Istrail 1997, Agarwala
{\it et al.} 1997). These models have had only a limited success in
understanding the energy landscape of the protein folding process.
My analysis suggests that the use of a diamond lattice and more details
of amino acid properties will bring the lattice models closer to reality.
Of course, if the preferences of amino acid sequences for certain lattice
positions are known (e.g. from the analysis of protein structure databases),
that can simplify the simulations further.

From the view-point of nanotechnology, construction of desired molecular
structures by chains that know how to fold is a conceptual shift from the
conventional approaches that use external sources (either to carve the
shapes or to assemble individual building blocks). To be able to do that,
the folding code of amino acid sequences must be deciphered; wide ranging
applications are obvious. One can also consider the simpler problem of
constructing two dimensional patterns by folding chains made of flat
building blocks. Such an approach will be a contrast to the standard
techniques of lithography, but it requires first figuring out the
appropriate two-dimensional building blocks and processes that can join
them together in chains. Carbon rings and the geometry of graphite sheets
will no doubt play a central role as the optimal ingredients in such an
exercise.

In closing, I want to contrast different information processing paradigms.
Electronic computation uses physical building blocks and operations based
on real variables. Quantum computation extends the building blocks and
operations to the complex numbers. Structural information processing goes
still one step further, to the non-commutative algebra of quaternions.
Systematic analysis of structural information processing has a long way to go.
Yet in a sense, it came first---proteins arose before genes, nervous signals,
spoken and written languages, number systems and computers. After all, the
word ``protein'' derives from the Greek ``protos'' meaning ``first'' or
``foremost''.

\section{Acknowledgements}

I am grateful to C. Ramakrishnan and M.A. Viswamitra for useful discussions
regarding protein structure and folding. The Ramachandran map in Fig.1 was
kindly provided by C. Ramakrishnan.

\section{Appendix: Properties of proteins}

Many careful experiments have been performed over the years to study various
properties of proteins. Here I list some of the important features they have
revealed, as a quick reference (Lehninger {\it et al.} 1993, Creighton 1992,
Fersht 1999):\\
$\bullet$ Proteins are synthesised by ribosomes by joining amino acids
together as a linear chain. The linear chain gradually folds into the
three-dimensional shape unique to every protein. The folding occurs by
rigid body transformations of the bonds---deformations such as
stretching/shrinking/bending of bonds are insignificant.\\
$\bullet$ Proteins carry out their tasks by binding to various molecules.
The binding is highly specific, very much like a lock and key arrangement.
Structurally stable features, precisely located on the protein surface,
are necessary for this purpose. When a protein fails to fold into its
proper shape due to some error, it cannot carry out its task and that gives
rise to a disease. (Sickle-cell anaemia was the first genetic disease to be
understood in this manner. A single mutation substitutes glutamic acid with
valine at the sixth position in the amino acid sequence. The resultant
defective haemoglobin does not fold correctly and is unable to carry out
its function properly.)\\
$\bullet$ The sequence of amino acids encodes the structural information
of a protein. Fig.3 shows the structure of individual amino acids. Proteins
may include other important components, e.g iron in haemoglobin, but the
role of these other components is essentially chemical and not structural.\\
$\bullet$ The sequence of amino acids is obtained by translation from the
sequence of nucleotide bases in DNA. This translation is necessary because
the two languages serve two different purposes, and the purposes decide the
physical components for their realisations. According to the cell's need,
proteins are synthesised, transported to appropriate locations to participate
in biochemical reactions, and degraded at the end. The three-dimensional
shape of the protein plays a critical part in its reactivity. The double
helical structure of DNA, with nucleotide bases hidden inside, protects the
one-dimensional information until it is required. DNA replication is also
much less error-prone than protein synthesis.\\
$\bullet$ The physical separation between consecutive amino acids in
polypeptide chains and consecutive nucleotide bases in DNA is about the
same, approximately $3.5$\AA. It is not direct stereochemistry, therefore,
which is responsible for three nucleotide bases being mapped to one amino
acid in the genetic code. The non-overlapping triplet code is likely to
have arisen from the need to have a sufficient number of amino acids as the
required building blocks for the three-dimensional protein structure. Living
organisms had to then set up the complex machinery, involving tRNA as
adapters connecting nucleotide bases and amino acids, to carry out the
task of translation.\\
$\bullet$ Correct translation is ensured by the bilingual aminoacyl-tRNA
synthetases that attach amino acids to tRNA molecules with appropriate
anticodons. There may be several anticodons which map to a particular amino
acid, but there is only one aminoacyl-tRNA synthetase per amino acid which
carries out the many-to-one mapping. Once the tRNA molecules are properly
charged with amino acids, the ribosomes match the anticodons of tRNA with
codons of mRNA and construct the polypeptide chain.\\
$\bullet$ Proteins often have to cross membranes and cell walls after their
synthesis, since they often have to carry out their tasks at locations
other than their place of synthesis. Membranes and cell walls cannot afford
to have big holes (otherwise many molecules would leak), and that provides
an important reason why proteins are folded chains. During translocation,
proteins unfold to their chain form, cross the barrier through a small hole
and then fold again into their native three-dimensional form.\\
$\bullet$ The three-dimensional protein structure specified by the sequence
of amino acids is essentially unique. Small proteins fold on their own, but
many large proteins require help of molecular chaperons to fold. Globular
proteins that fold on their own can be melted by heat, and they regain their
native form upon cooling.\\
$\bullet$ Carbon and nitrogen atoms, joined by strong covalent bonds, form
the back-bone of the polypeptide chain. In this chain rigid peptide bonds
alternate with rotatable bonds of $C_\alpha$ atoms. The $C-N$ peptide bonds
have a double bond character; the nitrogen atom carries a positive charge
making its electronic behaviour similar to the tetravalent carbon atom.\\
$\bullet$ Different amino acids are distinguished from each other by their
$R-$groups, which are side chains attached to the $C_\alpha$ atoms. Amino
acid $R-$groups are of various types: polar, non-polar, aromatic, positively
and negatively charged. The interactions of these $R-$groups with each other
and with the ambient water molecules fix the orientations of the rotatable
$C_\alpha$ bonds. These interactions are weak, and easily influenced by the
pH of the ambient liquid and the temperature.\\
$\bullet$ Atoms in proteins are quite densely packed. In terms of the van der
Waals atomic size, packing fraction for proteins is in the range $0.70-0.78$,
compared to $0.74$ for closest packing of identical spheres. The packing
fraction of a diamond lattice is only $0.34$, and the side groups of a
polypeptide chain folded along a diamond lattice fill up the empty spaces.
Even then the packing density is high along the chain, while the amino acid
side groups are somewhat loosely packed. Small cavities are filled up by
water molecules, which fit into tetrahedral geometry nicely.\\
$\bullet$ The polypeptide chain is synthesised in the fully extended form,
corresponding to $\phi=180^\circ=\psi$ and ``trans'' configuration. Certain
domains of proteins start folding as soon as they are synthesised, indicating
that at least some of the folding rules are local.\\
$\bullet$ The folding process occurs in stages. Local domains fold first,
essentially due to weak bonds (hydrogen and van der Waals). This process
is dominated by local transformations, i.e. proper rotations of bonds of
$C_\alpha$ atoms, and forms well-known structures such as $\alpha-$helices
and $\beta-$sheets. In the next stage, already folded domains get linked
by long-distance connections, e.g. disulfide bonds. In the final stage,
various separately assembled structures, polypeptide chains and chemical
groups, join together.\\
$\bullet$ Regular structures like $\alpha-$helices and $\beta-$sheets are
largely determined by the properties of the polypeptide back-bone, with a lot
of freedom in choice of amino acid $R-$groups. It is the irregular twists and
turns of the chain which critically depend on the interactions of the amino
acid $R-$groups. Several different type of interactions exist (electric
charges, dipoles, hydrogen bonds, rings, bifurcation and bulk of side chains
etc.) to achieve all possible shapes.\\
$\bullet$ In reality, proteins fold rather rapidly. The folded chain is a
self-avoiding walk in three-dimensional space. Such a walk can get stuck for
topological reasons, or it may need global criteria to complete its task
(traveling salesman type of problems are NP-hard with just local rules).
An easy escape is to complete the task with multiple walks, i.e. start a new
walk when the previous one gets stuck. Indeed many proteins are made of not
a single polypeptide chain, but several polypeptide chains entangled together.
Large protein structures are often made of polypeptide units arranged in
regular patterns (e.g. fibroin, keratin, collagen, virus coats etc.).\\
$\bullet$ Fig.2c shows the rotation angles around single bonds in a polypeptide
chain. Because of steric conflict between various atoms, not all the values
of angles occur in a polypeptide chain. The Ramachandran map ($\phi-\psi$
angular distribution) in Fig.1 displays the orientations available to the
amino acids. Experimental data for polypeptide chains follow the constraints
of the Ramachandran map quite well; in fact, the map is often used as a filter
for models when the experimental data are not accurate enough. The side chain
angles ($\chi$) also have preferred orientations which are separated by
$120^\circ$.\\
$\bullet$ Structural roles played by some of the amino acids are well-known.
Glycine with no side chain and no chiral centre is the most flexible. Proline
with its rigid ring and trans-cis transformation plays an important role in
forming sharp bends. Cysteine connects far separated regions of the polypeptide
chain by strong disulfide bonds, helping the folded chain retain its shape.\\
$\bullet$ Ordinary chemical reactions produce a mixture of D-type and L-type
molecules, but biological processes utilise molecules of only L-type chirality.
The smallest amino acid glycine is achiral, the next smallest alanine is once
in a while found in D-type configuration, while the rest are always in L-type
configuration. There exist racemase enzymes which can flip chirality of amino
acids. Most of the time they convert D-type amino acids to L-type for use in
protein synthesis. After a cell dies, its molecules gradually revert to a
mixture of D- and L-types. The proportion of D- and L-type molecules in a dead
cell can indeed be used to figure out how long back the cell died.

\bigskip
\centerline{\bf References}
\medskip

{\small
\noindent
Agarwala R, Batzoglou S, Dan{\v c}ik V, Decatur S E, Hannenhalli S, Farach M,
Muthukrishnan S and Skiena S 1997 Local rules for protein folding on a
triangular lattice and generalized hydrophobicity in the HP model;
{\it J. Comput. Biol.} {\bf 4} 275-296

\noindent
Arnez J G and Moras D 1997 Structural and functional considerations of the
aminoacylation reaction; {\it Trends Biochem. Sci.} {\bf 22} 211-216


\noindent
Creighton T E (ed) 1992 {\it Protein Folding} (New York: W H Freeman)

\noindent
Crick F H C 1966 Codon-anticodon pairing: The wobble hypothesis;
{\it J. Mol. Biol.} {\bf 19} 548-555

\noindent
Crick F H C 1968 The origin of the genetic code;
{\it J. Mol. Biol.} {\bf 38} 367-379

\noindent
Eriani G, Delarue M, Poch O, Gangloff J and Moras D 1990
Partition of tRNA synthetases into two classes based on mutually exclusive
sets of sequence motifs; {\it Nature} {\bf 347} 203-206

\noindent
Fersht A 1999 {\it Structure and mechanism in protein science: A guide to
enzyme catalysis and protein folding} (New York: W H Freeman)

\noindent
Grover L 1997 Quantum mechanics helps in searching for a needle in a haystack;
{\it Phys. Rev. Lett.} {\bf 79} 325-328 [quant-ph/9706033]

\noindent
Hart W E and Istrail S 1997 Lattice and off-lattice side chain models of
protein folding: Linear time structure prediction better than 86\% of optimal;
{\it J. Comput. Biol.} {\bf 4} 241-259

\noindent
Ikehara K 2002 Origins of gene, genetic code, protein and life: Comprehensive
view of life systems from a GNC-SNS primitive genetic code hypothesis;
{\it J. Biosci.} {\bf 27} 165-186

\noindent
Kolaskar A S and Ramabrahmam V 1982 Obligatory amino acids in primitive
proteins; {\it Biosystems} {\bf 15} 105-109

\noindent
Lehninger A L, Nelson D L and Cox M M 1993 {\it Principles of Biochemistry}
second edition (USA: Worth Publishers)

\noindent
Lewin B 2000 {\it Genes VII} (Oxford: Oxford University Press)


\noindent
Patel A 2000 Quantum algorithms and the genetic code;
{\it Proc. Winter Institute on Foundations of Quantum Theory and Quantum
Optics} (Calcutta), {\it Pram{\=a}{\d n}a} {\bf 56} 367-381
[quant-ph/0002037]

\noindent
Patel A 2001 Quantum database search can do without sorting;
{\it Phys. Rev. A} {\bf 64} 034303 [quant-ph/0012149]

\noindent
Ramachandran G N, Ramakrishnan C and Sasisekharan V 1963
Stereochemistry of polypeptide chain configurations;
{\it J. Mol. Biol.} {\bf 7} 95-99
}

\medskip\centerline{\small MS received 1 April 2002; accepted 2 May 2002}
\medskip\leftline{\small Corresponding editor: VIDYANAND NANJUNDIAH}

\end{document}